\documentclass[rpm,preprint,graphicx, onecolumn]{revtex4-1}
\usepackage{graphicx}% Include figure files
\usepackage{dcolumn}% Align table columns on decimal point
\usepackage{bm}% bold math
\usepackage{epstopdf} %converting to PDF
\usepackage [autostyle, english = american]{csquotes}
\usepackage{float}
\usepackage{color}
\usepackage{amsmath}
\usepackage[normalem]{ulem}
\usepackage{multirow,booktabs}
\MakeOuterQuote{"}
\begin{document}
%\preprint{APS/123-QED}

\title{\textbf{Grain Boundary Resistance in Copper Interconnects \\
from an Atomistic Model to a Neural Network }}% Force line breaks with 

%\maketitle
% Force line breaks with 
%\thanks{A footnote to the article title}%

\author{Daniel Valencia}
\email{valencid@purdue.edu}

\author{Evan Wilson}

\affiliation{Birck Nanotechnology Center, Network for Computational Technology,
Purdue University, West Lafayette, 47907, USA }%

\author{Zhengping Jiang}

\affiliation{ Samsung Semiconductor Inc, San Jose CA, 95134, USA }%

%\altaffiliation[Also at ]{Samsung Semiconductor Inc, San Jose CA, 95134, USA.}

\author{Gustavo A. Valencia-Zapata}

\author{Gerhard Klimeck}

\author{Michael Povolotskyi}

\affiliation{ Birck Nanotechnology Center, Network for Computational Technology,
Purdue University, West Lafayette, 47907, USA }%

%\date{\today}% It is always \today, today,
%  but any date may be explicitly specified
\begin{abstract}
Orientation effects on the  {\color{black}  specific resistance} of copper grain boundaries are studied  systematically with two different  atomistic tight binding methods.
A methodology is developed to model the  {\color{black} specific resistance}  of grain  boundaries {\color{black} in the ballistic limit} using  the Embedded Atom Model, tight binding  methods and non-equilibrum Green's functions (NEGF).  The methodology is validated against first principles calculations for   {\color{black} thin films with a single coincident grain boundary},  
with  6.4\% deviation in the  {\color{black} specific resistance}. A statistical ensemble of 600 large, random  structures with grains is studied.  
For structures with three grains, it is found that  the distribution of  {\color{black} specific resistances} is  close to  normal. Finally, a compact model 
for grain boundary  {\color{black} specific resistance} is constructed based on a neural network. 
%\begin{description}
%\item[Usage]
%Secondary publications and information retrieval purposes.
%\item[PACS numbers]
%May be entered using the \verb+\pacs{#1}+ command.
%\item[Structure]
%You may use the \texttt{description} environment to structure your abstract;
%use the optional argument of the \verb+\item+ command to give the category of each item. 
%\end{description}
\end{abstract}
%\pacs{Valid PACS appear here}% PACS, the Physics and Astronomy                            % Classification Scheme.
%\keywords{Suggested keywords}%Use showkeys class option if keyword
%display desired
\maketitle
%\tableofcontents
\section{\label{sec:level1} INTRODUCTION}
Due to the aggressive downscaling of logic devices, interconnects
have reached the nanoscale, making quantum effects important. 
According to the roadmap provided by ITRS, interconnects are expected to reach
sizes of 10 to 30 nm in the next decade \cite{ITRS:2014:Online}. Previous
work by Graham et al. \cite{Graham2010} demonstrates that surface
scattering and grain boundary (GB) scattering play major roles in the  {\color{black} resistance} 
of structures smaller than 50 nm.  
Earlier works based on semi-empirical parameters have
described polycrystalline films and surface scattering \cite{Fuchs1938,Mayadas1969}
for macroscopic systems, but   {\color{black} the fact that those models require fitting parameters for each experimental setup} limits the
scope of their applications. The ultra-scaled interconnects suggested
by the roadmap require better descriptions of orientation and confinement
effects to correctly model scattering in wires. Recently, first-principles calculations have been used to describe the  {\color{black} 
resistance of a single grain boundary} by making use of non-equilibrium Green's function
with Density Functional Theory (DFT-NEGF) formalism \cite{Cesar2014}. The results demonstrate a strong
correlation between  {\color{black} resistance} and {the geometry of the \color{black} grain boundary }, and show 
agreement with both experimental \cite{Kim2010} and other theoretical
work \cite{Feldman2010,Zhou2010,Zhang2007}. However, the studied structures
are limited to relatively small sizes containing single grain boundaries
and less than a few hundred atoms because of the computational burden
required to perform DFT-NEGF calculations.  \\
{\color{black} The purpose of this manuscript is to introduce an atomistic model 
that describes the resistivity due to grain boundary effects 
for realistic copper interconnects as projected by the ITRS roadmap~\cite{ITRS:2014:Online}  
without depending on any phenomenological parameter. Even though the atomistic model is much faster
than an \textit{ab initio} method, parametric models have the advantage of easily providing 
a quantitative value of resistivity. Therefore, a compact model  which reduces the computation time is  generated by making use of a neural network that is based on large statistical sample. The rest of the manuscript has been organized as follows. 
Section ~\ref{sectionII} presents the main characteristics of the atomistic models
and benchmarks tight binding parameters against
first principles calculations for a copper FCC structure. Section ~\ref{sectionIII}
constructs single grain boundaries based on coincident site lattice (CSL)
and validates their electronic properties against an \textit{ab initio} method. 
Section \ref{sectionIV} describes grain boundary effects on copper interconnects using a system of three grains of 10 nm length simulated with an atomistic
method which is benchmarked in the previous sections and quantifies the effect of misorientation.
Section \ref{sectionV} proposes a compact model based on three different algorithms and finds that a neural network approach
 best matches the results obtained from the atomistic methods, allowing the results to be generalized to any grain boundary system configuration with a total length of 30 nm.
Section \ref{sectionVI} presents a summary of this work.
}

%\newpage
\section{\label{sectionII} Description of Tight Binding Models}
The two tight binding methods used in this study 
are an environmental orthogonal tight binding model (TB) \cite{Hegde2014} and
a non-orthogonal tight binding method based on the Extended H\"{u}ckel
(EH) model \cite{Hoffmann1988}. The TB model has 
an  orthogonal basis with an interaction radius up to the second nearest neighbor (2NN). However,
it requires a large number of parameters to include strain effects (48 parameters for copper). In comparison, the EH model has a 
non-orthogonal basis with a larger interaction radius up the third nearest neighbor (3NN). 
It requires a smaller number of parameters than the TB method (11 parameters for copper). 

\begin{figure}[ht]
\includegraphics[width=6cm,height=11cm,keepaspectratio]{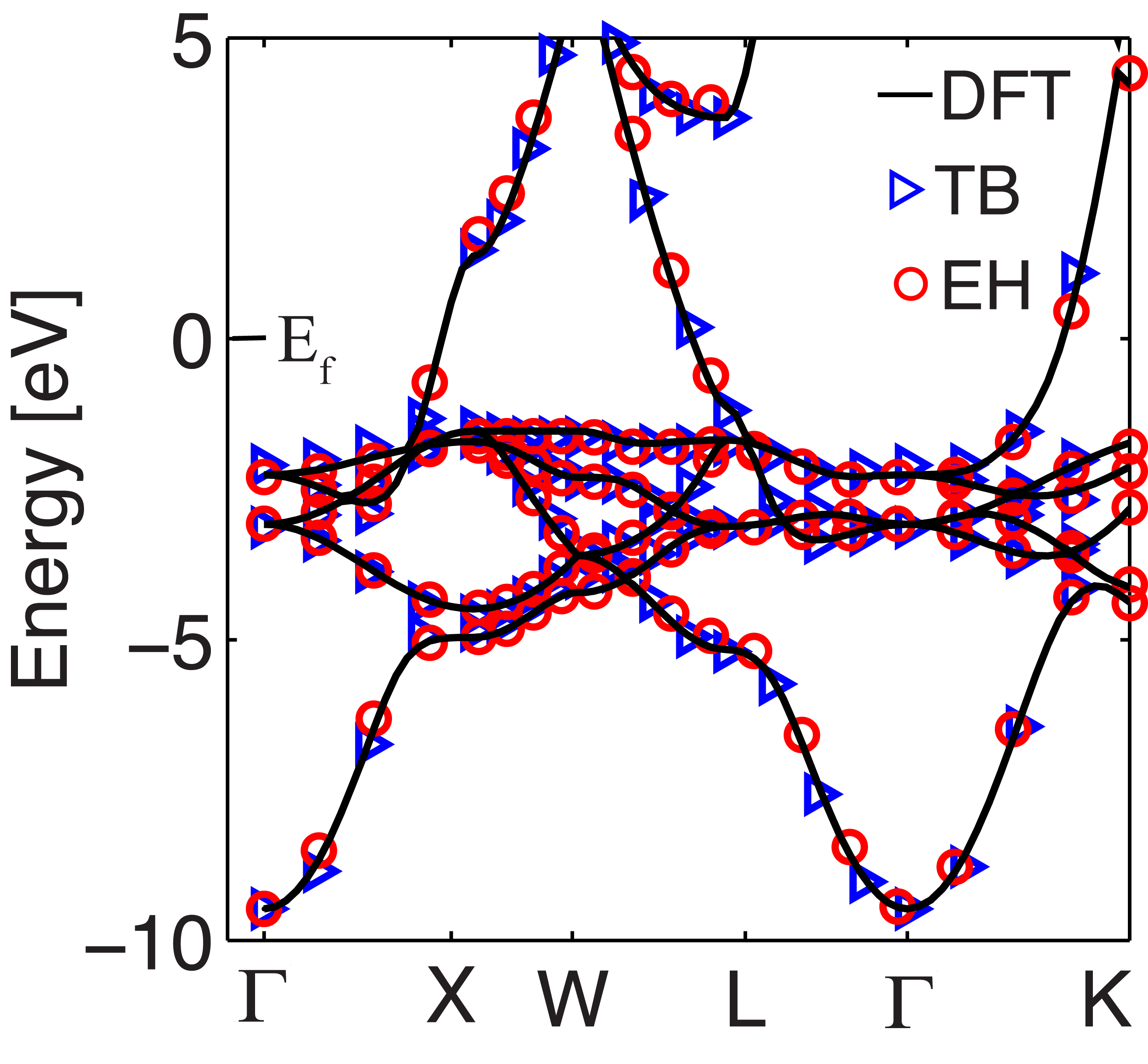}
\caption{Band structure for copper unit cell obtained by TB, EH  
and DFT methods. $\rm E_f$ indicates the Fermi energy.}
\label{Cu_band} 
\end{figure}

Existing parameters for the TB model~\cite{Hegde2014} fail when used in highly distorted atomic systems such as GB.
Due to the exponential dependence of the inter-atomic coupling on the bond length, the inter-atomic matrix elements corresponding to
bond lengths with a 5\% or greater distortion generate unphysical results. The problem is solved
by obtaining a new parametrization with additional constraints on the inter-atomic coupling. This new parameter set is summarized in  TABLE~\ref{TB_table}  in Appendix A.  The parameters for the EH model are taken from literature \cite{Cerda2000}.  Both EH parameters and the new TB parameters show a good match for the Cu unit cell when compared  against an \textit{ab initio} method {\color{black} as  shown in Fig. \ref{Cu_band}}. The  \textit{ab initio} result, used as a reference, is obtained by density functional method with a
Perdew-Burke-Ernzerhof version of the
generalized gradient approximation (GGA PBE) exchange- correlation functional~\cite{Brandbyge2002}.  An energy cutoff
of 150 Ry is used  and the Brillouin zone is sampled 
with a  10$\times$10$\times$10 mesh. An FCC copper lattice with a lattice constant of 0.361 nm, as reported
experimentally \cite{Straumanis1969}, is considered.
\newpage
\section{\label{sectionIII} Coincident Site Lattice Grain Boundaries}
\begin{figure}[ht]
\includegraphics[width=7cm,height=3.25cm]{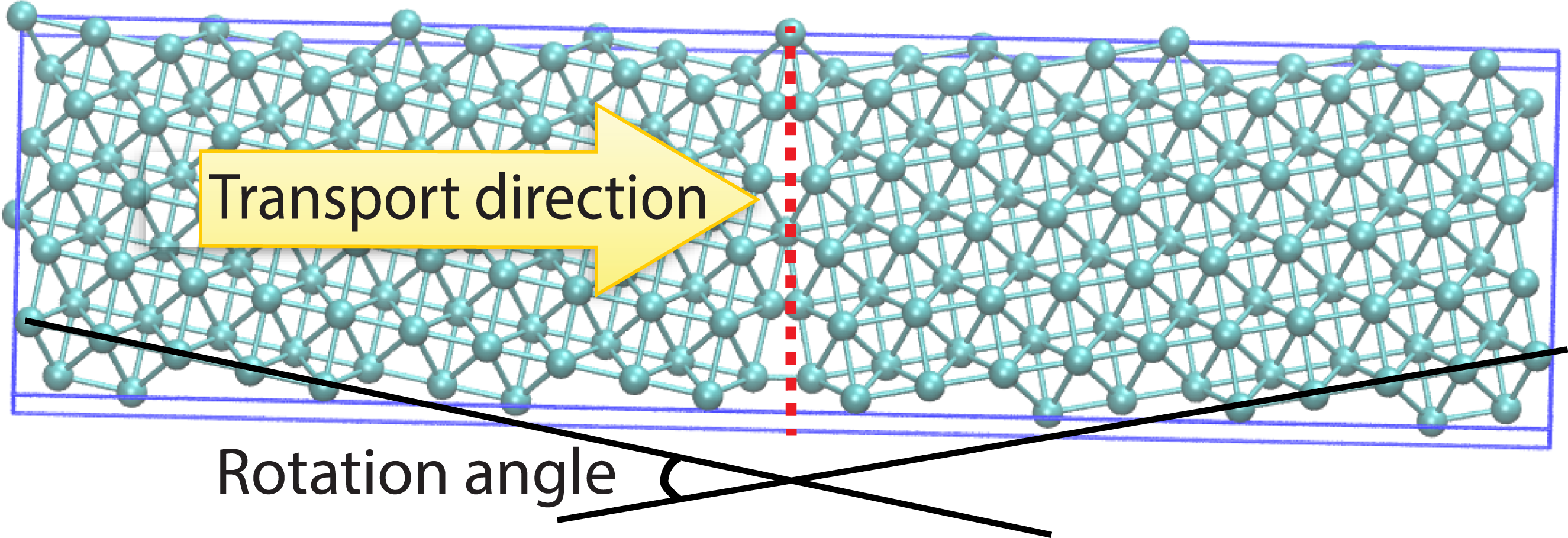}
\caption{Coincident site lattice GB are obtained by generating a superposition of two periodic lattices.
One of the lattices was rotated with respect to the other, generating
coincident points between the lattices for each rotation angle.}
\label{CSL_lattice}
\end{figure}

To validate the tight binding models, the effects of GB scattering 
were studied for a single coincident site lattice grain boundary.
The simplest GB configurations are obtained by a rotation of one of the grains 
until its lattice vector becomes coincident with the vector of the unrotated lattice ~\cite{CSL_ref} as shown in Fig.~\ref{CSL_lattice}.
A   fairly small number of atoms ($<400$) is required to construct these systems, which allows the
tight binding models to be benchmarked against a first principles calculation as implemented in the ATK package \cite{Brandbyge2002}. 

CSLs are labeled by $\Sigma N$, where $N$ corresponds to the ratio of the CSL
unit cell size to the standard unit cell size. In this work, the CSL GB are generated
with GBSTUDIO~\cite{GBSTUDIO:2014:Online} and  relaxed using an \textit{ab initio} method.
The relaxation is carried out with GGA PBE  exchange-correlation functional.  A Double Zeta polarized basis set 
is used for copper atoms with an energy cutoff is 150 Ry and the Brillouin
zone sampled with a  4$\times$4$\times$1 mesh, until all atomic forces on each ion are
less than $10^{-5}$ eV/{{\AA}}.
Once the ionic relaxation is completed, the transmission spectrum for  CSL structures
is calculated by the recursive Green's function method \cite{Lake1997} implemented in NEMO5 \cite{Steiger2011}
in an energy range between -2 and 2 eV around the Fermi level  with a   Brillouin
zone sampled  with a  30$\times$30$\times$1 mesh. 
\newpage 

\begin{figure}[ht]
\includegraphics[width=0.55\textwidth]{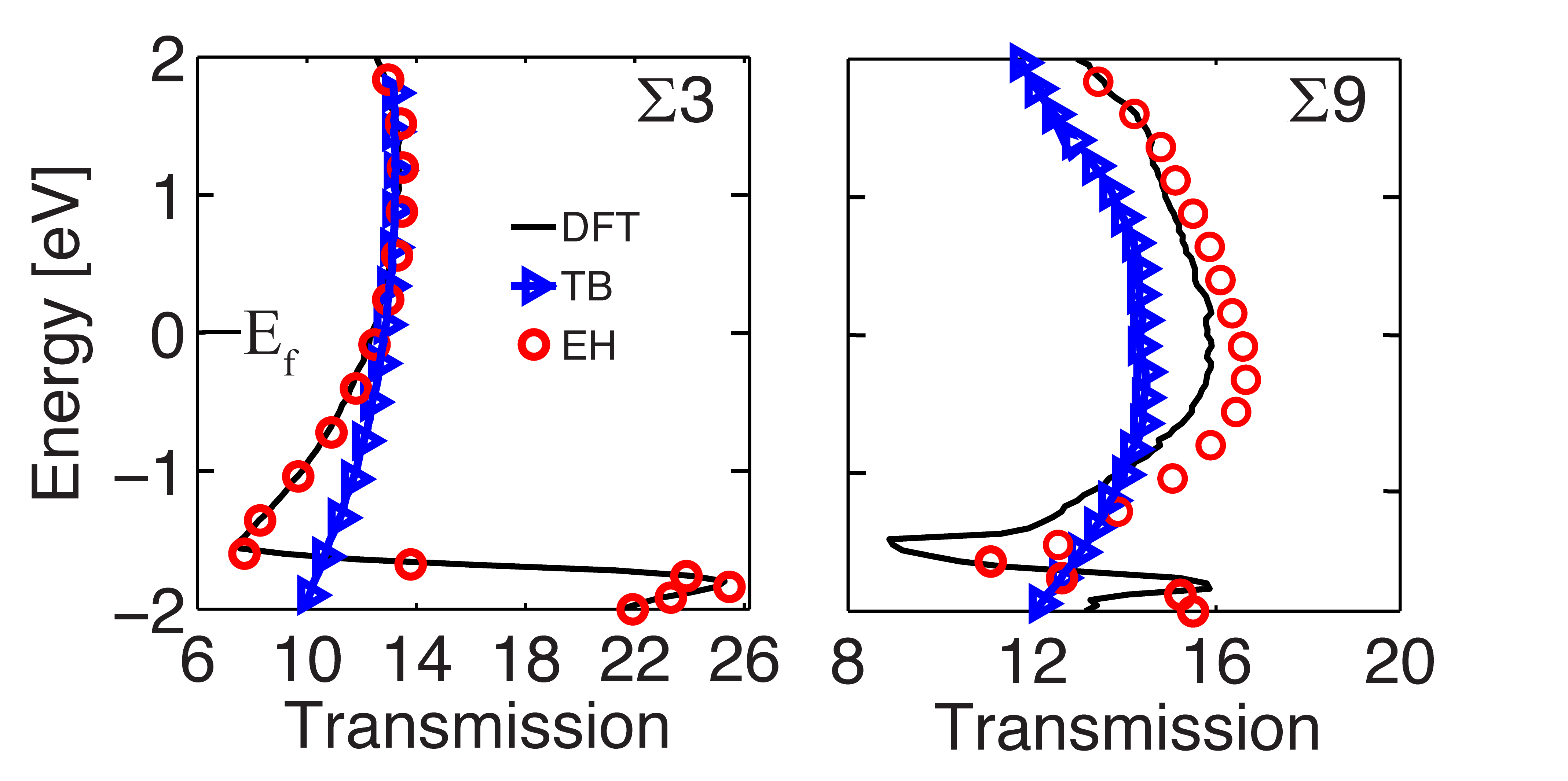}
\caption{Transmission spectra $T(E)$ for two different CSL ($\Sigma3$ and $\Sigma9$)
show that EH captures the main features of DFT.}
\label{S5_S9_Trans} 
\end{figure}

The integrated  transmission spectra in the k space obtained by the  tight binding methods are compared against the spectrum obtained by the \textit{ab initio} method with a similar basis set,  energy cutoff 
and Brillouin mesh as is used in the ionic relaxation. 
The results in Fig.~\ref{S5_S9_Trans} show that the EH method captures the main features of DFT not only at the Fermi energy ($E_f$), but also over
a large energy window. On other hand, while the transmission spectrum calculated
by TB also shows reasonable agreement with DFT around the
Fermi window, it fails to describe the \textit{ab initio} transmission spectrum for energies away from
the $E_f$.
 \\
Subsequently, the resistance for the CSL GB {\color{black} in the ballistic limit} is obtained based on the Landauer formalism assuming
a low bias condition \cite{SupriyoDatta1997}  as: 
\begin{eqnarray}\label{single_eq}
G=\frac{1}{R}= \frac{2\, e^{2}}{h} \int T(E_{f},{\bf k})\,d^2k, 
\end{eqnarray}
where $G$ is the conductance, $R$ is the resistance, $e$ is the
elementary charge, $h$ is Planck's constant and $T(E_{f},{\bf k})$ is
the transmission for a particular  wave vector ${\bf k}$  at the Fermi energy. 
The Fermi levels in Figs.~\ref{Cu_band} and~\ref{S5_S9_Trans} are calculated at the leads of the device self consistently for DFT and non-self consistently for tight binding models. In this last case, the Fermi level  is obtained by integrating over the DOS from $-\infty$ to $E_f$ until this  value becomes equal to the total number of electrons at a zero  temperature approximation \cite{Valencia2016}. 
Following Ref.~\cite{Cesar2014} the  {\color{black} specific resistances}  of the CSL grain boundaries are obtained by 
 {\color{black}$\gamma^R = (R-R_B)\, A$, where
$R$ is the resistance of the configuration that contains the GB, $R_B$ is the resistance of the perfect bulk copper, and	
A is the grain cross section.
The specific resistances for those CSL configurations are calculated by TB and EH and compared to DFT} as shown in  Table~\ref{CSL_resistance} 

\begin{figure}[ht]
\includegraphics[width=0.5\textwidth]{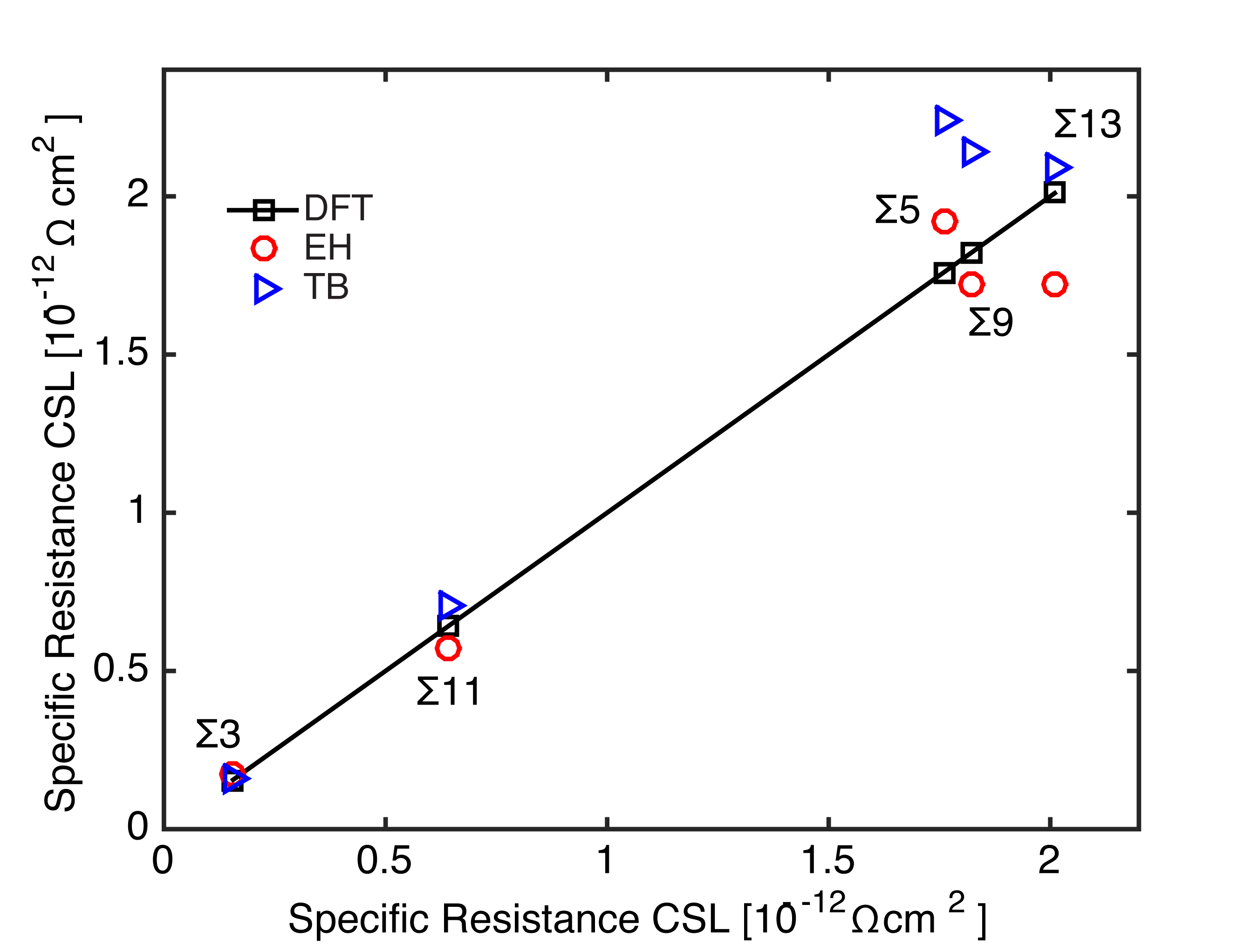}
\caption{Resistivities for different CSLs, labeled  by $\Sigma N$, calculated by TB and EH and compared with the DFT method.}
\label{resistance_CSL} 
\end{figure}

\begin{table}[h]
\centering
\begin{tabular}{c c c c c c}
\hline\hline
%\cline{1-6}
%\cmidrule{1-2}\morecmidrules\cmidrule{1-2}
&\multicolumn{5}{c}{Specific resistance CSL $\gamma^R \, (10^{-12} \Omega\, cm^2 )$} \\
\cline{2-6}
%&\multicolumn{2}{c|}{High information density}&\multicolumn{2}{c|}{Low information density}&\\ 
%\hline
\midrule
CSL GB &  $\gamma_{DFT}$  &   $\gamma_{EH}$   &   $\gamma_{TB}$  & Experimental & Other References \\
\cline{1-6}
%\hline
$\Sigma3$    &0.156&0.173 & 0.158 &0.170 ~\cite{Lu2004} &0.202 ~\cite{Kim2010}   \\ [-0.2cm]
             &     &      &       &                     &0.155 ~\cite{Zhou2010}  \\ [-0.2cm]
             &     &      &       &                     &0.158 ~\cite{Cesar2014}  \\ [-0.2cm]
             &     &      &       &                     &0.148 ~\cite{Zhang2007}   \\
%\hline
$\Sigma5$    &1.759& 1.934& 2.240 &						& 1.885 ~\cite{Kim2010}  \\ [-0.2cm]
             &     &      &       &                     & 1.49 $\,$ ~\cite{Cesar2014}   \\
             
$\Sigma9$    &1.82 & 1.72 & 2.14  & 					& 1.75 $\,$ ~\cite{Cesar2014} \\
%\hline
$\Sigma11$   &0.64 & 0.57 & 0.71  &                     & 0.75 $\,$	~\cite{Cesar2014}		  \\  
%\hline 
$\Sigma13$&2.01 & 1.72 & 2.09  	  &                     & 2.41 $\,$	~\cite{Cesar2014}	  \\  
\hline \hline 
\end{tabular}
\caption{Specific resistance for  different CSL ($\Sigma N$) calculated by TB, EH and DFT. }
\label{CSL_resistance}
\end{table}
%\newpage
The results {\color{black} in the Table~\ref{CSL_resistance} and Fig.~\ref{resistance_CSL} }  show
less than 10.4 $\%$ difference in the {\color{black} specific resistance} between EH and DFT, and less than
$11.2\%$ between TB and DFT.  Thus the atomistic methods (TB and EH) are  able to describe copper
interconnects with reasonable accuracy. These methods are chosen to study GB systems with $10^3$ to $10^4$ atoms 
because they require significantly fewer computer resources than the \textit{ab initio} calculations~\cite{Valencia2016}.
 \\

%\begin{table}[H]
%\begin{center}
%\begin{tabular}{|c|c|c|c|}
%\hline 
%&$\rho$ DFT & $\rho$ TB & $\rho$ EH \\  
%&$(\Omega\, cm^2 10^{-12} )$ &$(\Omega\, cm^2 10^{-12})$ & $(\Omega\, cm^2 10^{-12})$ \\  \hline
%$\Sigma3$ &9.4& 9.6& 9.8 \\  \hline 
%$\Sigma5$ &10.4& 10.8& 11.0 \\  \hline 
%$\Sigma9$ &12.8& 13.3& 14.0 \\  \hline 
%$\Sigma11$ &9.6& 9.6& 9.9 \\  \hline 
%$\Sigma13$ &12.9& 13.7& 13.9\\  \hline 
%\end{tabular} 
%\caption{Resistivities ($\rho$) for different CSL ($\Sigma N$) calculated by
%TB, EH and DFT.}
%\label{resistance_table}
%\end{center}
%\end{table} 
Only non \textit{ab initio} methods are capable
of relaxing structures of this size ($\gg10^{3}$ atoms),
 {\color{black} therefore a force field potential method based on an} 
Embedded Atom Model (EAM)   is {\color{black} used.}    {\color{black}The relaxation is performed using LAMMPS software package}~\cite{Lammps}   {\color{black} with an EAM potential constructed by Y. Mishin  et al. that is fitted to first principles calculations to correctly describe grain boundaries and point defects in copper ~\cite{Mishin2001}. }.
\\~\\ %~\\~\\
The accuracy of this approach is determined by comparing the 
{\color{black} formation energy  %defined in Eq.~(\ref{Eq:Form_Ener})
for CSL GBs obtained} by \textit{ab initio} and the EAM method. The formation energy $\gamma^E$ is defined as follows:
 {\color{black} 
\begin{equation}\label{Eq:Form_Ener}
\gamma^E=\frac{E_{slab}-NE_{0}}{A},
\end{equation}
where $E_{slab}$ is the total energy of a slab configuration that contain a CSL GB, $N$ is the number of atoms in the CSL GB, $E_0$ is the energy of a single atom of  bulk copper and $A$ the cross sectional area.} The ionic relaxation carried out by \textit{ab initio} methods used the plane wave DFT package (VASP)~\cite{Kresse1996} 
and a PBE GGA  exchange-correlation functional. The plane wave energy cutoff is 500 eV
and the Brillouin zone is sampled with a  4$\times$4$\times$1  mesh, until all atomic forces on each ion are
less than $10^{-5}$ eV/{{\AA}}.   
%The \sout{total}  {\color{black} formation} energy ($\gamma^E$) comparison is \sout{tabulated in Table~\ref{energy_table} where $E_{DFT}$ and  $E_{EAM}$}  
%{\color{black} shown in Figure ~\ref{Fig:Formation_Energy}.
\begin{figure}[ht]
\includegraphics[width=0.6\textwidth]{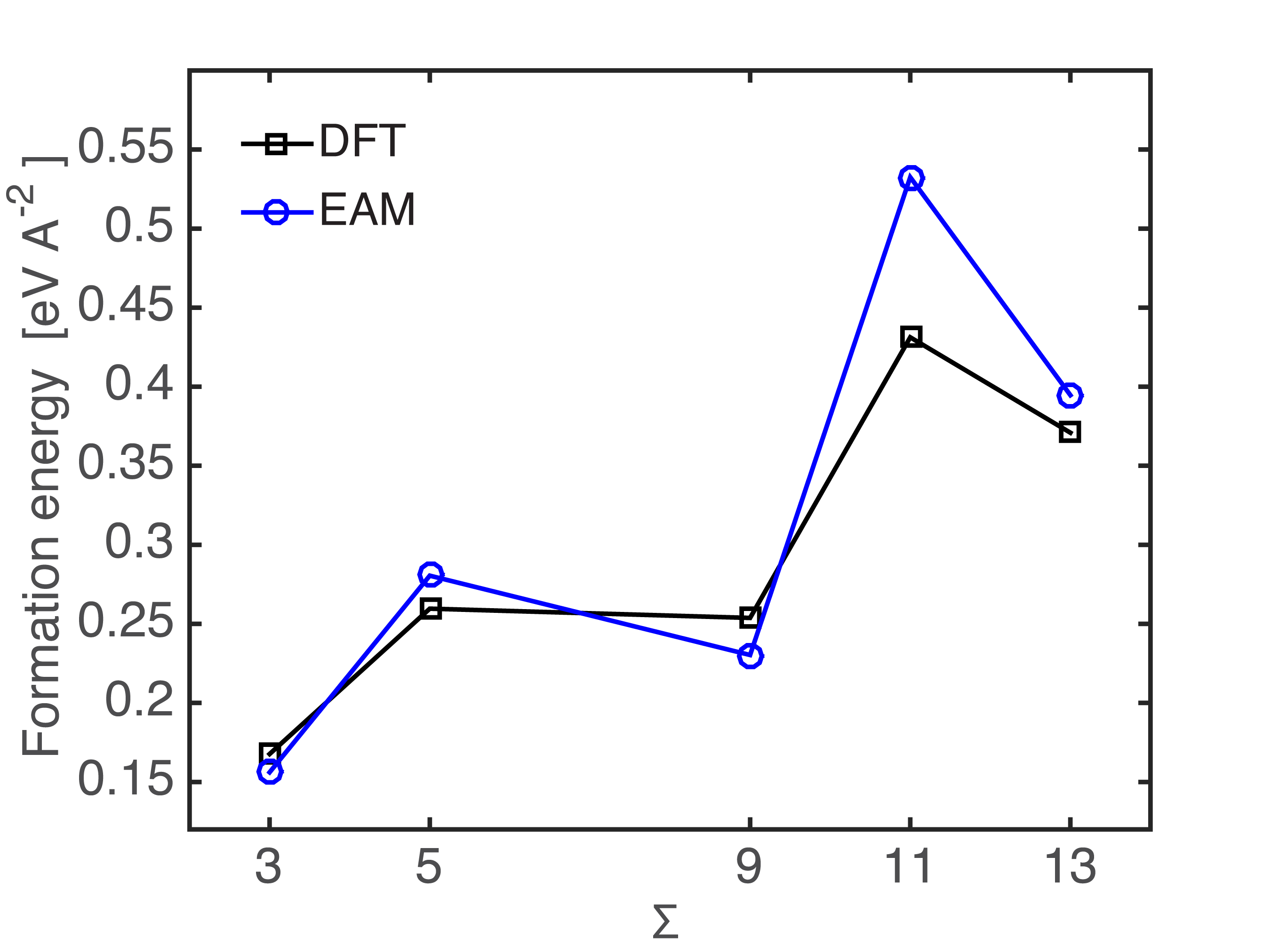}
\caption{Formation energy ($\gamma^E$) for different CSLs GB, labeled  by $\Sigma N$, relaxed by DFT and EAM potential.}
	\label{Fig:Formation_Energy} 
\end{figure} 
Comparison of the relaxation energy, computed using the EAM potential, with the DFT result (see Fig.~\ref{Fig:Formation_Energy}), shows that
the difference is less than  7\% with for all CSL orientations except the $\Sigma_{11}$,  which shows a larger error of  20~\%. 
These results indicate that the EAM potential calculation is an acceptable method to relax the grain boundary structures with 
the benefit of reduced computational burden, compared to DFT.

\section{\label{sectionIV}  Specific Resistance for Grains of 10 nm length} 
\begin{figure}[ht]
	\includegraphics[width=0.8\textwidth]{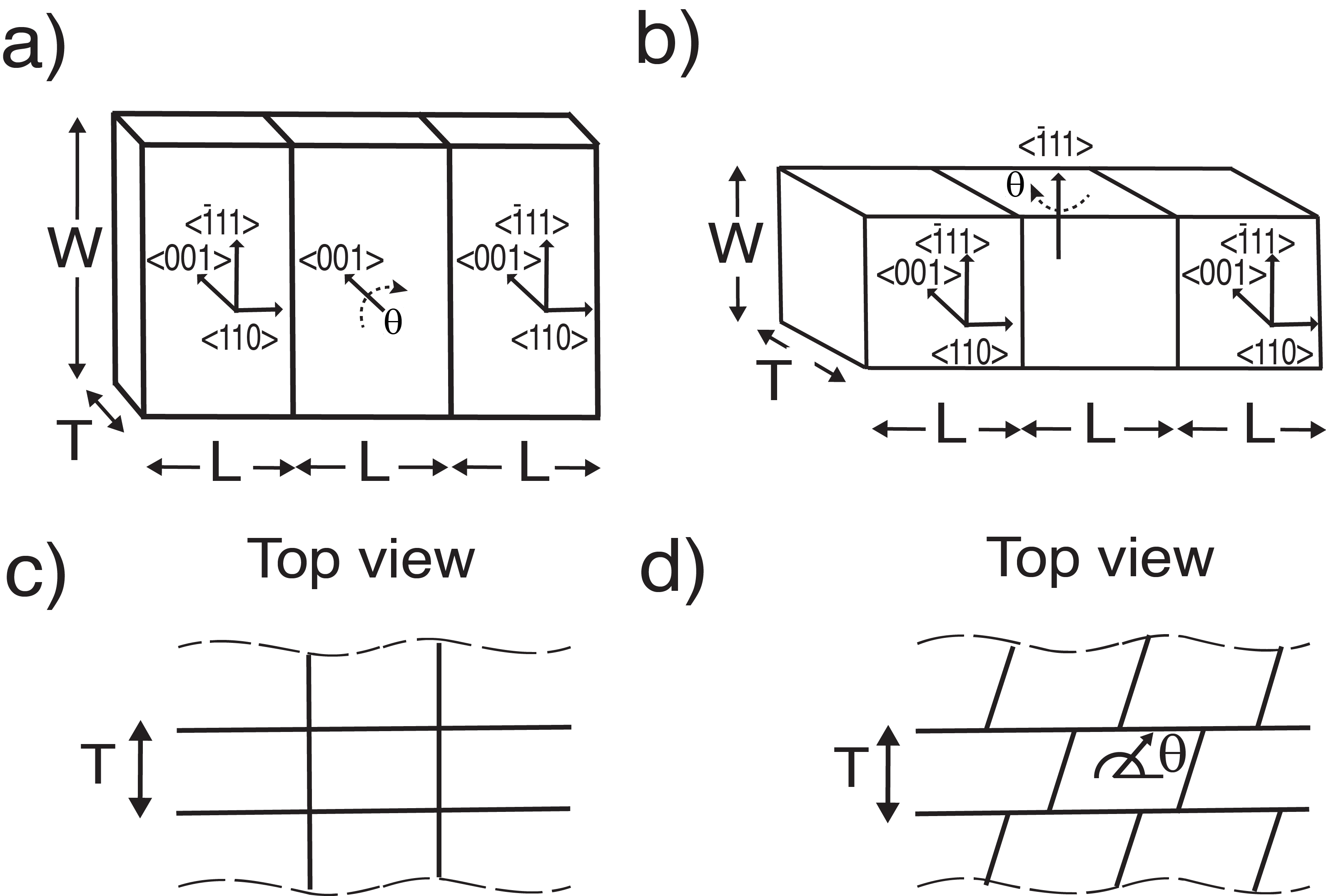}
	\caption{GB classification: a) "Tilt" GB are generated by a rotation in  the  {[}001{]}  plane 
		b) "Twist" GB generated by a rotation in the  {[}$\bar{1}$01{]}  plane {\color{black} where the grain boundary is always perpendicular to the transport direction}. c) and d) figures represent the top view of "Tilt" and "Twist" GB configurations. }
	\label{GB_struct} 
\end{figure}

Based on the prediction of the ITRS roadmap that interconnects will
reach 10 to 30 nm {\color{black} length} in the coming years \cite{ITRS:2014:Online}, {\color{black} a set of copper thin films of 30 nm  is constructed and modeled by tight binding methods as described in Section II. The copper interconnects are formed by three grains of 10 nm length.
Each grain is constructed with a super cell  growing in the  ${[}110{]}$  orientation with a lattice constant of 0.361 nm which has the highest conductance \cite{Hegde2014}, as reported experimentally \cite{Straumanis1969}.  In order to quantify the effect of GB orientation on the specific resistances  for copper interconnects},  two different types of GBs are generated by  Voronoi diagrams \cite{Rycroft2006}.  These GB types are  based on the rotation direction of the middle grain shown as  "Tilt" and  "Twist"  GBs  respectively, {\color{black} which generates two boundaries} as shown in
Fig.~\ref{GB_struct} a)  and b). {\color{black} In order to have a lower impact on the specific resistivity  due to the electrode setup, 
three grains are modeled in this work.}
%{\color{red} why three grains?  } \\
%\newpage
{\color{black}  In both configurations, only the middle GB is initially rotated then a periodic boundary condition is  applied in the ${[}001{]}$ direction for the ionic relaxation and the electronic transport calculation. Therefore,  atomic surface roughness  is present in the structures  as a result of the relaxation. Additionally it is assumed that each configuration shown in the Fig. 6 and 8 is connected to a   pristine source and drain lead oriented in the [110] direction, whose  atoms are fixed during the ionic relaxation.}
\\~\\
The  "Tilt"  GBs are  generated by a rotation {of the middle grain} with respect to the {\color{black}  {[}001{]}} direction 
by an angle $\theta$ in a range between 0 and $\pi/2$. {\color{black}  Each grain is formed by a supercell of 10 nm length ($L$) in the transport direction ${[}110{]}$, 10 nm width ($W$) in the ${[}\bar{1}01{]}$ direction  and 0.361 nm thickness ($T$) in the periodic direction  ${[}001{]}$ as shown in Fig.~\ref{GB_struct} a) and c)}. \\

The "Twist" GBs 
{\color{black} are generated by a rotation of the middle grain with respect to the {[}$\bar{1}$11{]} direction}
 by an angle $\theta$ in a range between 0
and $\pi/2$. 
{\color{black} The rotation is applied in the same direction as the periodicity, therefore thicker grains are constructed to ensure the grains overlap after rotation.} 
%\sout{In order to use similar computational resources as in the "Tilt" GBs, the width is decreased}.  
In this configuration setup  each grain is formed by a supercell of 10 nm length ($L$) in the transport direction ${[}110{]}$, 3 nm width ($W$) in the   ${[}\bar{1}01{]}$ direction and  3 nm thickness ($T$) in the periodic direction  {[}001{]} 
as shown in Fig.~\ref{GB_struct} b) and d). \\
It is important to clarify that after any rotation for "Tilt" or "Twist" GB the [110] direction is no longer the transport direction for that grain. Similarly, the rotation angle corresponds to the initial value, but this value will be slightly modified after relaxing the structure.  
%\newpage
 \\

\begin{figure}[ht]%[H]
	\begin{center}
		\includegraphics[width=0.5\textwidth]{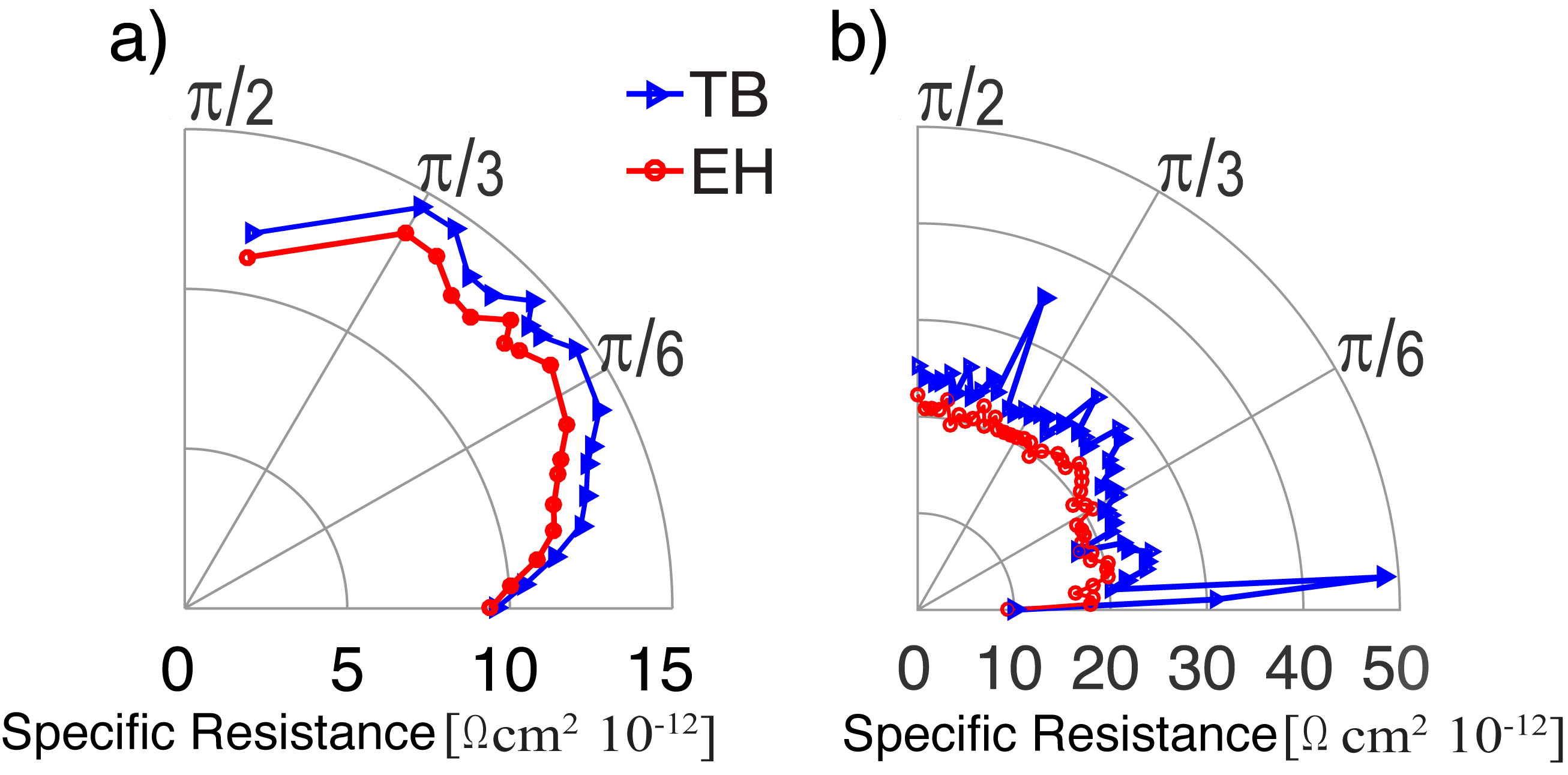}
		\caption{ a) "Tilt" and b) "Twist" GB  {\color{black} specific resistance} calculated by the TB and EH methods.}
		\label{GB_sphere} 
	\end{center}
\end{figure}

\begin{figure}[ht]
\includegraphics[width=0.5\textwidth]{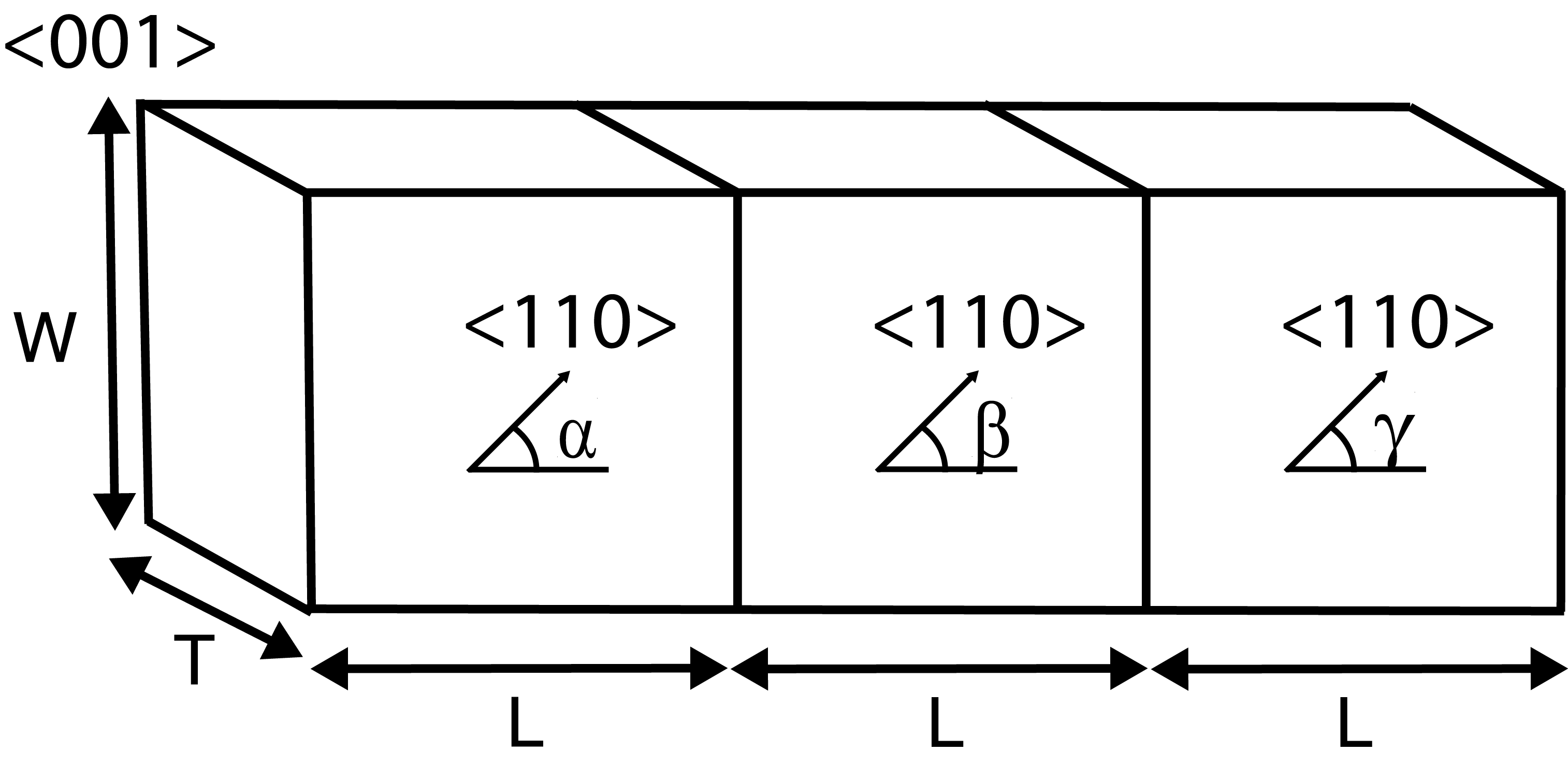}
\caption{ GB configuration constructed with three grains, each one  generated by rotating the lattice through angles   $\alpha, \beta, \gamma$, 
respectively, around the ${[}001{]}$ axis.} 
\label{GB_large_three} 
\end{figure}

%\newpage
The {\color{black} specific resistance} for "Tilt" and "Twist" GBs {\color{black}  for different orientations 
are obtained by a procedure similar to that described in Section III as $\rho=R\times A$,  where $R$ is obtained by Eq. (\ref{single_eq}) 
and each configuration is relaxed by an EAM potential. In order to compare the specific resistivity for "Tilt" and "Twist" GBs for different angles $\theta$, the  "Tilt" GBs values are normalized such that "Tilt" and "Twist" GBs are calculated over the same cross sectional area. } Those values 
are plotted in Fig.~\ref{GB_sphere}.  In both systems, {\color{black} specific resistance} increases with an increase in the angle, until the angle reaches $\pi/6$, and then  
becomes almost constant, although the "Tilt" GB shows a
reduction after $\pi/3$. The {\color{black} specific resistance} dependence for  {"Twist"} GBs shows more noise than for {"Tilt"} GBs, because
"Twist"  structure has more points per unit area where the grain boundaries intersect  (see Figs.~\ref{GB_struct}c, d), which leads to a higher number
of dislocations. Differences between TB and EH, especially pronounced for "Twist" GBs, are due to the fact that the TB model does not correctly describe strained systems, where
the atoms are coupled by distances much smaller than the bulk bond length.  In particular, for the "Twist" system  with  rotation angles such as  6, 8 and 70 degrees, 
there are many atoms with a small distance between nearest neighbors which results in unphysical peaks in the{\color{black} specific resistance} dependence (see Fig.~\ref{GB_sphere}b) when it is calculated by the TB method.  
  \\
%\newpage
In order to understand and create a compact model to predict how  {\color{black} specific resistance} changes as a result of GB orientation,
a sample set of 600 configurations  are generated. Each GB is  constructed with three grains
and  each of them is rotated with an angle ($\alpha, \beta, \gamma$) in a range between 0 to 180 degrees parallel to the GB boundary. 
The dimensions of the GB are similar to those used for  "Tilt" GB  with thickness, width and length 
equal to 0.5 nm, 3 nm and  10 nm respectively as  shown in Fig.~\ref{GB_large_three}.
A periodic boundary condition in the {[}001{]} direction is imposed.
%\newpage
 
%\newpage
\begin{figure}[h] %[H]
\begin{center}	
\includegraphics[scale=0.3]{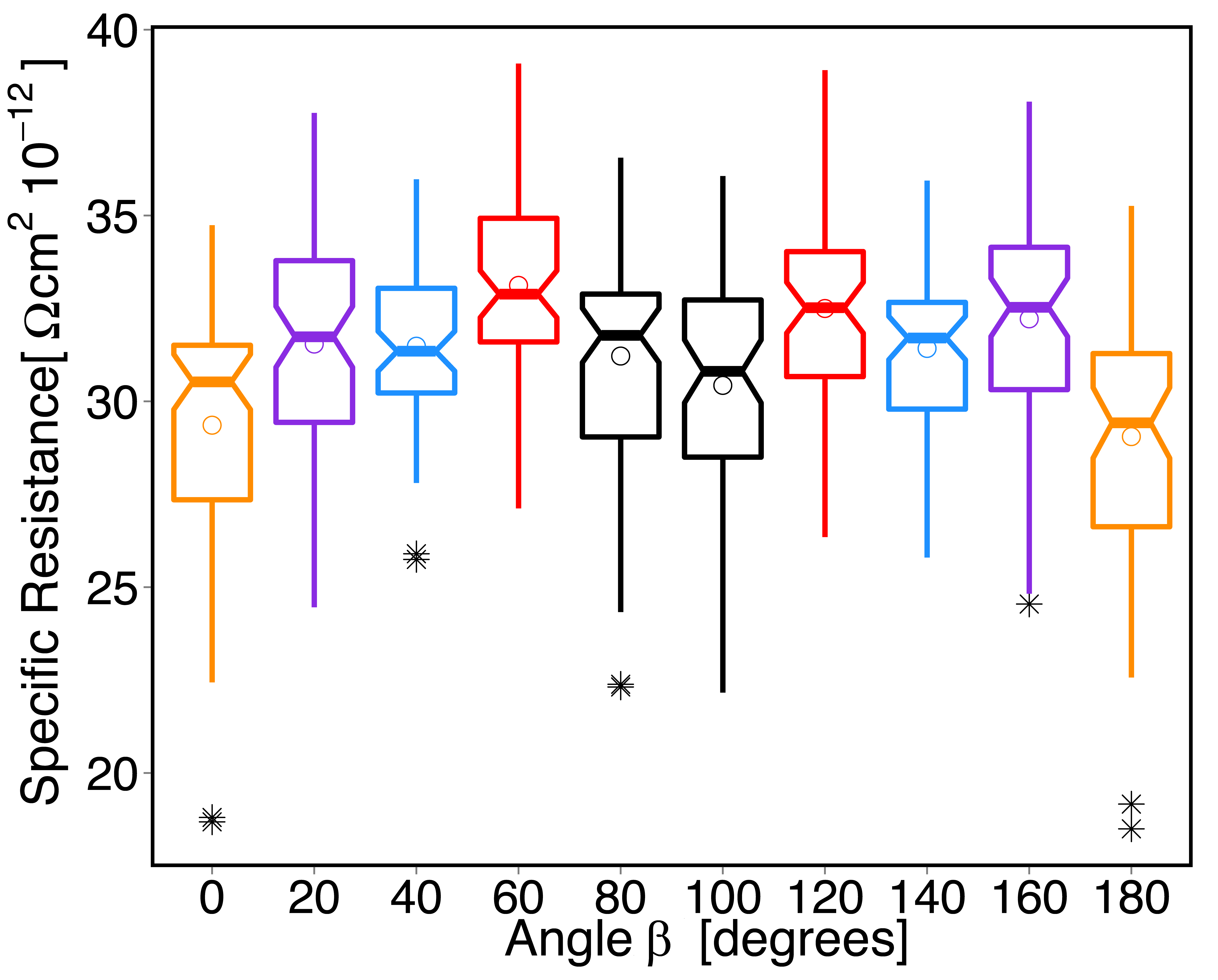}
\caption{Resistivity distributions for {\color{black} $\alpha$ and $\gamma$ between $0$ to $180$ degrees  and a constant angle $\beta$}. The boxplots represent the resistance distribution, while those marked with a star represent
outliers.}
\label{boxplot_large} 
\end{center}
\end{figure}

\begin{figure}[h]
\includegraphics[scale=0.5]{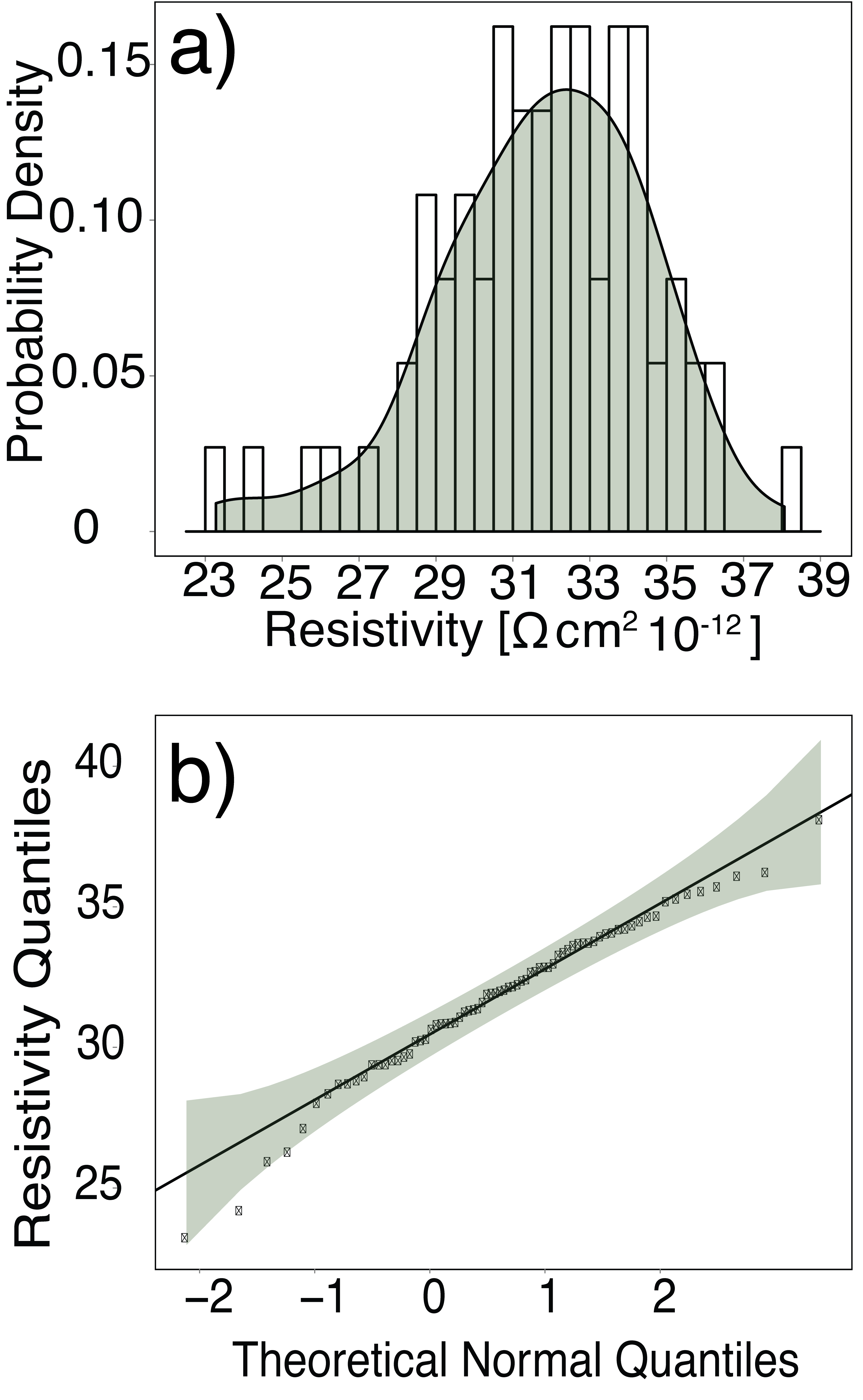}
\caption{ a) Probability distribution for a GB system rotated over three different angles $\alpha$, $\beta$ and $\gamma$ in a range between 90 and 180 degrees.
The shaded area represents the best approximation of a normal distribution for the 600 samples; b) Q-Q plot which confirms the normal distribution.}
\label{Distribution} 
\end{figure}

%\newpage
The  {\color{black} specific resistance} for these samples is calculated with the EH method because it is more reliable over angle rotations than the TB method. Making use of the results obtained from these samples, a boxplot for
{\color{black}  $\alpha$ and $\gamma$ in a range between $0$ to $180$ degrees}  and a constant angle $\beta$ is plotted in Fig.~\ref{boxplot_large} which shows a symmetry 
in the  {\color{black} specific resistance} in a range between 0 to 90 degrees 
and 90 to 180 degrees. This observation  is  confirmed by a statistic nonparametric Kolmogorov-Smirnov test~\cite{Conover1999}
which compares the distribution function for the group of samples in a range between 0 to 90 degrees against those between 90 to 180 degrees and  finds that both
groups of samples  are drawn from  an equivalent, continuous distribution. A p-value  of  0.16 is obtained for the Kolmogorov-Smirnov test,  confirming that there is no difference
between the {\color{black} specific resistance} distributions for both cases with a confidence of 95$\%$. The symmetry in the  {\color{black} specific resistance} is due to the fact that the crystal symmetry of copper
is not totally disrupted by the
structural relaxation. The probability distribution for the three different angles ($\alpha$, $\beta$ and $\gamma$) in a range between 90 to 180 degrees is plotted in Fig.~\ref{Distribution}. 
%\newpage
%\newpage
Per the Shapiro-Wilk test \cite{Conover1999} with a p value of 0.15 and a 95\% confidence, the  {\color{black} specific resistance} distribution follows a normal distribution with a mean
and standard deviation equal to 31.7 $\times 10^{-12}\, \Omega\,cm^2$ and  2.8 $\times 10^{-12}\, \Omega\,cm^2$. 
The Q-Q plot in Fig.~\ref{Distribution} b) 
shows that the  {\color{black} specific resistance} distribution is likely normal, although the left and right tails do not follow a normal distribution. 
%\clearpage
%\vfill
 
%\newpage
\section{\label{sectionV} Grain boundaries Modeled by a Neural Network}
Atomistic models based on a tight binding approach can describe the effects of the GB orientation on  {\color{black} specific resistance} with the same accuracy as DFT methods, but with a much lower computational burden.
However, {\color{black} the specific resistance calculated by } atomistic models such as  EH and TB {\color{black}  }for a combination of three grains of 10 {\lowercase{nm}} length in the transport direction  are still not as fast as  {\color{black} conventional models
such as the Fuchs-Sondheimer and  Mayadas-Shatzkes models ~\cite{Fuchs1938,Mayadas1969}	which describe surface roughness and  grain boundary effects
respectively in  copper interconnects. However these models require experimental input to fit some  parameters which limits the transferability for different configurations. Therefore,  compact models based on the statistical results obtained from an atomistic model described in Section~\ref{sectionIV} are proposed to  describe the scattering effects on grain boundaries for a system of 3 grains of 10 {\lowercase{nm}} length.  Three different algorithms are used to construct the compact models, including a polynomial fit, a nearest neighbor search model and a neural network as described in the following subsections. The inputs for the compact models are the orientation angles $\alpha$, $\beta$ and $\gamma$ and the output is the
specific resistance  of the GB $\rho(\alpha,\beta,\gamma)$. The compact models are trained with a random selection of 80\% of the 600 samples plotted in the Fig.~\ref{Distribution} and validated with the remaining 20\% of the data.} \\~\\
\subsection*{Polynomial Fit}
{\color{black}
A polynomial fit of second order  is carried out  based on a least squares adjustment, obtaining the following parametric relationship between the misorientation angles $(\alpha,\beta,\gamma)$ and the  specific resistivity: }
\begin{align} 
\rho(\alpha,\beta,\gamma) =&  21.95 + 10.59\alpha - 2.76\alpha^3 + 10.54\beta \\ \nonumber
                           &- 6.15\beta^2 + 13.41 \gamma - 3.91 \beta \gamma - 5.18\gamma^2
\end{align} \\

\begin{figure}[h]
\includegraphics[width=0.5\textwidth]{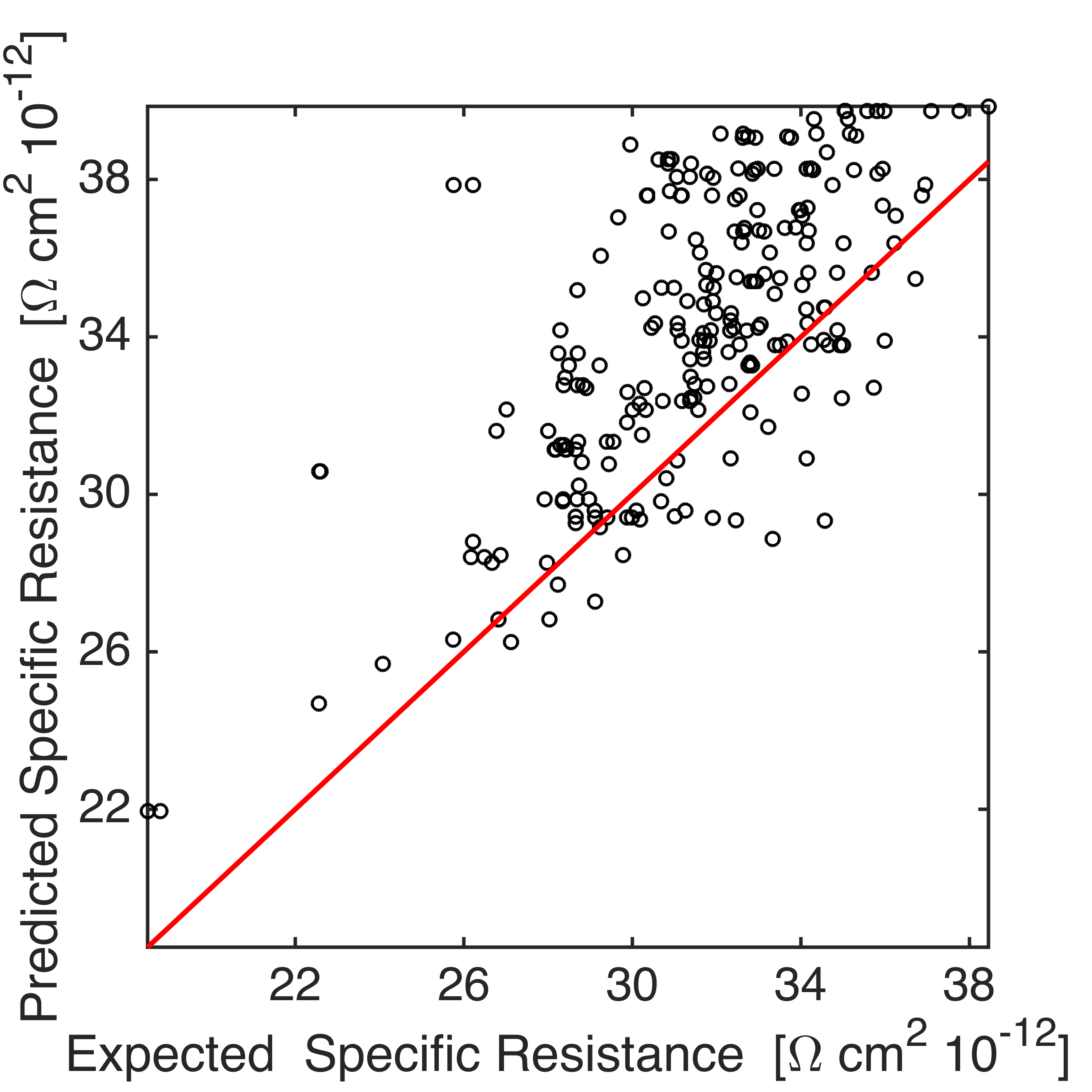} 
\caption{Evaluation of the  specific resistance for the multivariate polynomial model using least squares adjustment  
for the remaining 20\% of $\rho(\alpha,\beta,\gamma)$ values for copper interconnects.}
\label{Fig:Linear_fit} 
\end{figure}

{\color{black}
The expected values obtained from the model are compared against the remaining  20\% of the atomistic data as show in the figure ~\ref{Fig:Linear_fit}. The parametric fitting based on a polynomial approximation displays a poor match with the atomistic results with a 70\% variability of the resistivity for the training dataset and a MSE equal to 13.94 $\times 10^{-12}\, \Omega\,cm^2$. This result  shows that grain boundary effects cannot be modeled as a simple additive effect between each orientation. Therefore, a more complicated dependency exists between the resistivity and the orientation angles.}
   
\subsection*{Nearest Neighbor Fitting}
\begin{figure}[h]
\includegraphics[width=0.5\textwidth]{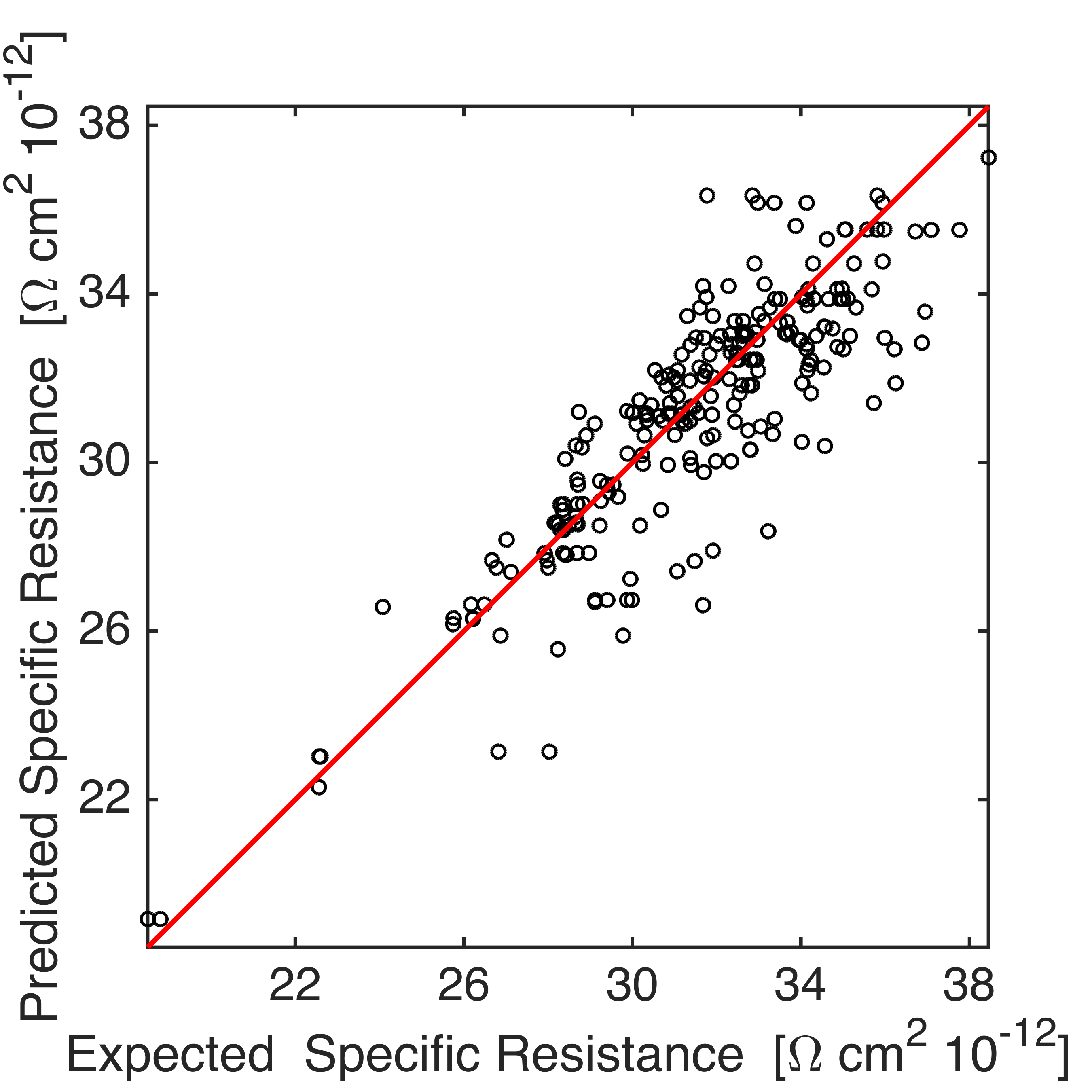} 
\caption{Evaluation of the  specific resistance for the Nearest Neighbor model for the remaining 20\% of $\rho(\alpha,\beta,\gamma)$ values for copper interconnects.}
	\label{Fig:Nearest_fit} 
\end{figure}

{\color{black} Since the polynomial fit provides a poor fitting for the  specific resistance of a GB  oriented by the angles $(\alpha,\beta,\gamma)$, a  non-parametric model is explored based on a "Nearest-Neighbor" search which uses the  
"dsearchn" triangulation method implemented in Matlab's optimization package \cite{Matlab}.
The comparison between the expected specific resistance and the predicted specific resistance
obtained from the testing data is plotted in Fig.~\ref{Fig:Nearest_fit}.  While this algorithm exhibits a mean square error  for the specific resistance  equal to 2.67 $\times 10^{-12}\, \Omega\,cm^2$ which is much lower than the error of the polynomial method, it does not support systems that have more than 3 degrees of freedom.}
%the authors should compare the predictive performance of the model with the much simpler (brute force) fitting
%of resistance vs angular space. Based on Fig. 7, the resistance vs angle seems to be a relatively smooth function
%
%Performs linear interpolation of 3dimensional data $\alpha$, $\beta$ and $\gamma$  scattered data described 
%by the points X and values V. The interpolation is based on a n-dimensional 
%delaunayn triangulation. A barycentric interpolation scheme is employed for 
%all query points using  dsearchn. 
%it can only handle  2D and 3D scatter data
%the authors should compare the predictive performance of the model with the much simpler (brute force) fitting
%of resistance vs angular space. Based on Fig. 7, the resistance vs angle seems to be a relatively smooth function

%\newpage
\subsection*{Neural Network Model}
\begin{figure}[h]
\begin{center}
\includegraphics[width=1.0\textwidth]{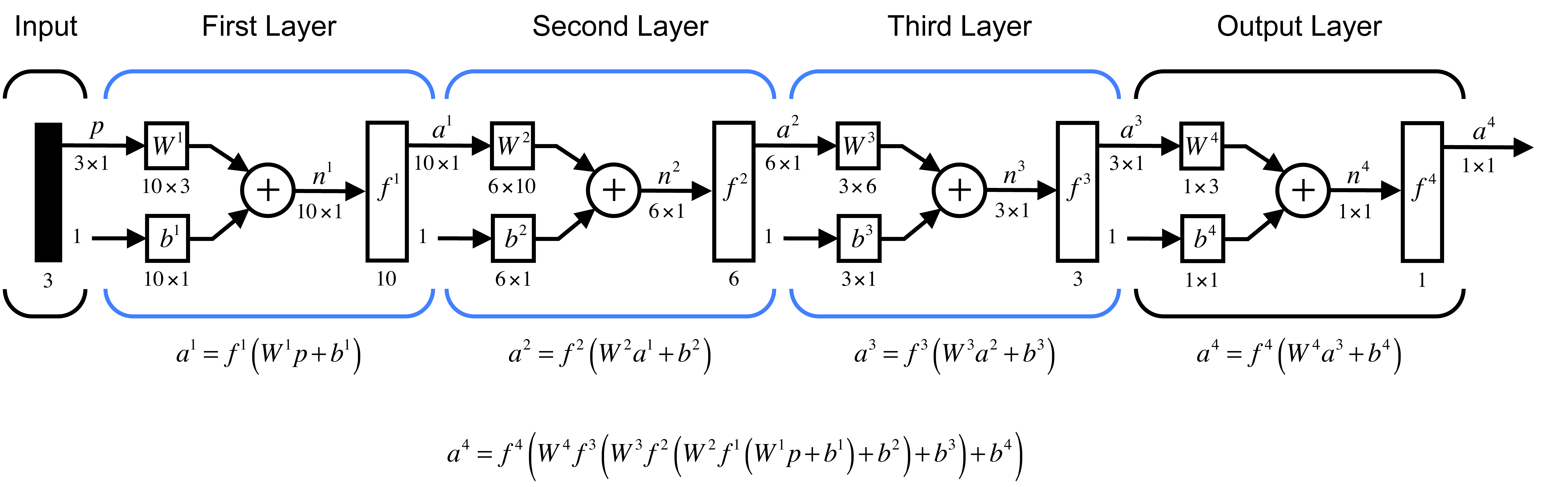}
\caption{Schematic representation for the Multi-Layer Neural Network used to describe grain boundary  {\color{black} specific resistance} for copper interconnects with three grains. 
The values \textbf{\textit{W}}$^i$ and \textbf{\textit{b}}$^i$ correspond to the weights and bias parameters, 
\textbf{\textit{f}}$^i$  represents  logistic functions  except for the last layer \textbf{\textit{f}}$^4$  to which is applied a linear 
function and  \textbf{\textit{a}}$^i$ corresponds to the output at each neuron $i$.  } \label{MLN_scketch} 
\end{center}
\end{figure}

%\clearpage
%\vfill
{\color{black} Finally, } a compact model based on a Neural Network (NN) \cite{NeuronalBook} algorithm is introduced.  NN models have been widely used to model complex problems; in the TB approach, NN algorithms  have been used to describe potential minimization \cite{NNPot} and materials parametrization \cite{NNTB}.  In this work,  a multilayer neuronal network (MLN)  is applied with a back-propagation algorithm \cite{NeuronalBook} to 
quickly obtain the  {\color{black} specific resistance} of the GB.  . \\~\\
The neural network shown in Fig.~\ref{MLN_scketch} is achieved  after testing different types of neural networks and varying  the number of 
hidden layers. The final system is formed by an input layer, three hidden layers, and one output layer.  The input layer \textbf{\textit{p}}$=(\alpha,\beta,\gamma$) is represented by
a  row vector of  dimension $3\times1$. The hidden layer is composed of  three inner layers $i$ with  10, 6, and 3 neurons, respectively; the 
weight \textbf{\textit{W}}$^i$ and   bias \textbf{\textit{b}}$^i$  vectors for a given layer $i$ are
shown in Fig.~\ref{MLN_scketch}. 

The MLN is implemented in the statistical software R making use of the package Neuralnet \cite{Rpack}. 
The value of the parameters \textbf{\textit{W}}$^i$ and   bias \textbf{\textit{b}}$^i$  are obtained by the gradient descent method \cite{NN_Descent}  
which  minimizes the mean square error of the output layer. In the NN, the functions \textbf{\textit{f}}$^i$  represent 
logistic functions employed at  each layer, except for the last layer \textbf{\textit{f}}$^4$  to which is applied a linear function. \\
\onecolumngrid
%%\twocolumn
%\newpage

\begin{figure}[h]
	\includegraphics[width=0.5\textwidth]{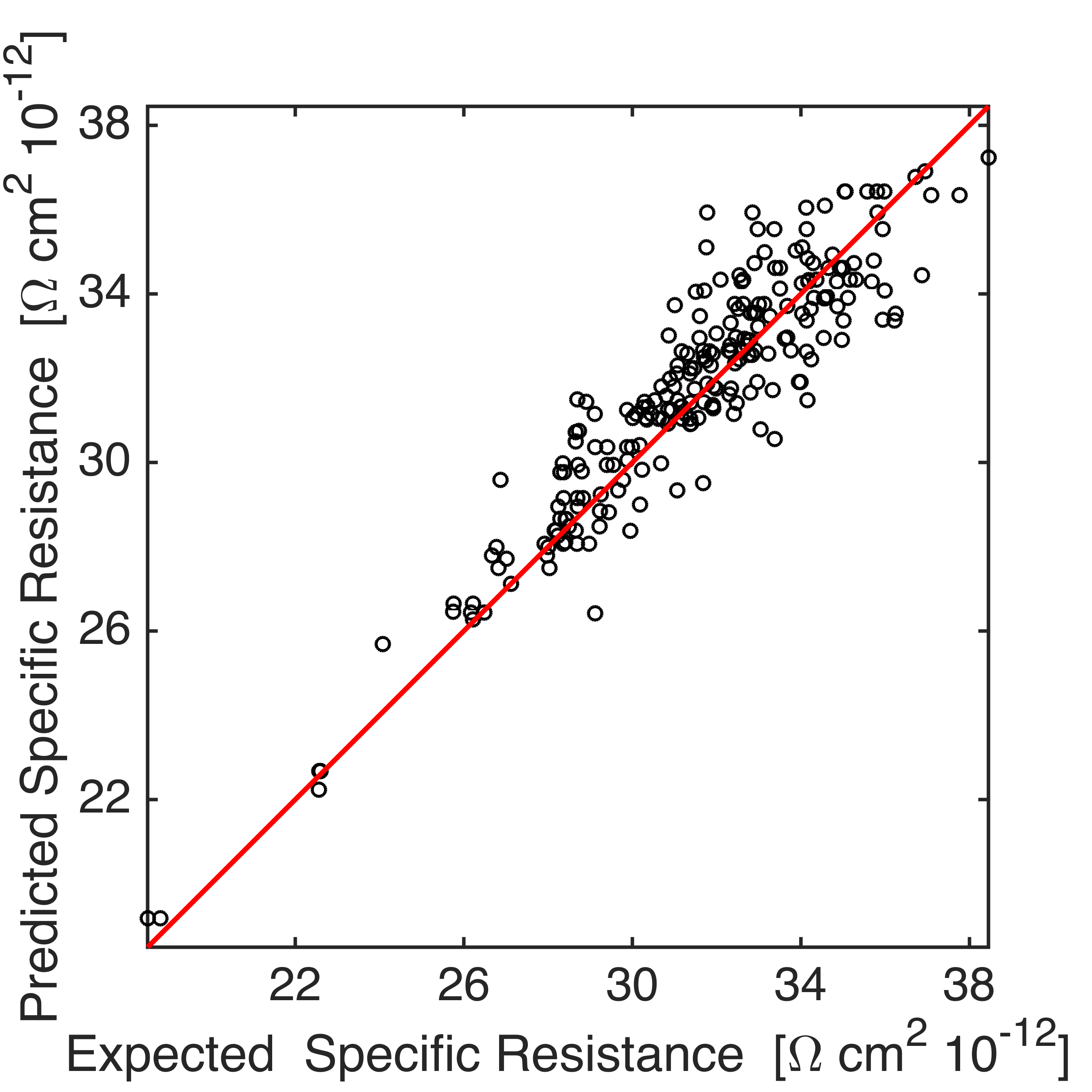}
	\caption{Evaluation of the  specific resistance for the Multi-Layer Neural Network model for the remaining 20\% of $\rho(\alpha,\beta,\gamma)$ values for copper interconnects.}
	\label{Neuron_expectValues} 
\end{figure}

%\twocolumngrid
%\newpage
The mean square error (MSE) prediction for  {\color{black} specific resistance} for this NN is  
 {\color{black} 1.44}$\times 10^{-12}\, \Omega\,cm^2$
.
The results obtained for the testing data of the MLN are plotted in Fig.~\ref{Neuron_expectValues};
the model shows good agreement for low values of  {\color{black} specific resistance}
and larger variability for GB with a  {\color{black} specific resistance} over the range 29.0 - 39.0$\times 10^{-12}\, \Omega\,cm^2$.
 \\
{\color{black} Comparing the three methods described above,  it is observed  that the neural network method  has a much smaller MSE than the other methods.  It can also be generalized to describe more complicated configurations with different geometries and a number of grains not possible with non-parametric methods such as "Nearest Neighbor" or linear fitting.}
%\newpage
\section{\label{sectionVI} SUMMARY}
In summary, the effect of orientation on grain boundary resistance for copper interconnects is studied using two different atomistic tight binding methods (EH and TB).
The transmission spectrum and  {\color{black} specific resistance} calculated by these methods are benchmarked for coincident site lattice single GB  ($\Sigma N$) against first principles calculations.
These results   show that the EH method captures the main features of DFT in the Fermi window between -2 to 2 eV. On other hand, the transmission spectrum calculated 
by TB also shows reasonable agreement with DFT around the
Fermi window, but fails to describe the \textit{ab initio} transmission spectrum  for energies away from
the Fermi energy. Since the computational requirements for tight binding methods are also much smaller than for first principle calculations, 
the EH method is an effective way to describe the  {\color{black} specific resistance} of 
 {\color{black} interconnects with lengths greater than  30 nm.}  

Orientation effects for  "Tilt"   and  {"Twist"}  GBs 
for copper interconnects of 30 nm length relaxed by a semi-classical EAM potential  are also benchmarked against first principles. Rotations perpendicular  
to the transport direction have a larger effect on the  {\color{black} specific resistance} of the GB than rotations parallel to the transport direction. 
Statistical analysis of GB  {\color{black} specific resistance} shows that the inversion symmetry of copper is still manifested for the considered grain  geometry.

Finally,  statistical models  based on {\color{black} three different algorithms are studied. The parametric model based on a polynomial fit of the misorientation angles $(\alpha,\beta,\gamma)$ shows a poor match with the test results from the atomistic model, 
confirming that a complex relationship exists between the specific resistance and the orientation angles.  While the nearest neighbor model displays a better fit with an error of 2.67 $\times 10^{-12}\, \Omega\,cm^2$ , it can only support three degrees of freedom. Among the studied models, the compact model based on neural network is the best algorithm to describe the specific resistance with a MSE lower than 1.44 $\times 10^{-12}\, \Omega\,cm^2$.  Additionally, the neural network can be used for systems with more than three degrees of freedom.} 

{\color{black} In this manuscript, the ballistic resistivity due to the grain boundary effect has been studied. While electron phonon scattering are reported to play an important role in copper resistivity  at room temperature and when the grains are  larger ~\cite{Graham2010,Plombon2006}, these effects have not been included in this work.  Future work will use the neural network to generate a compact model that includes electron-phonon scattering in addition to grain boundary effects to describe the resistivity for copper interconnects.}

\section*{ACKNOWLEDGMENTS}
This work was supported by the FAME Center, one of six centres of STARnet, a Semiconductor Research Corporation program sponsored by MARCO and DARPA. Support by the US Department of Energy National Nuclear Security Administration under Grant No. DE-FC52-08NA28617 is acknowledged. The authors also acknowledge the staff and computing resources of both the Rosen Center for Advanced Computing (RCAC) at Purdue University and the
Blue Waters sustained-petascale computing project, which is supported by the National Science Foundation
(award number ACI 1238993). Finally, the authors would like to thank  Dr. Bozidar Novakovic and David Guzman  
for stimulating discussions about the topic.
\subsection{\label{sec:level6} Appendix}
Parameters for bulk copper with the environmental tight binding method (TB) 
are obtained by direct fitting bulk band structure~\cite{Hegde2014}, but additional constraints on the inter-atomic coupling are included during the parametrization process.
%\begin{center}
\begin{table}[h]
{\scriptsize{
\begin{tabular}{|c|c|c|c|}  \hline 
Parameter Name  & Value & Parameter Name  & Value \tabularnewline \hline 
		$VBO$  & 3.6540 &$I\_D\_D\_\Delta$ & -0.08 \tabularnewline \hline 
		$E\_S$  & -4.5236  &$q\_D\_D\_\sigma$ & 4.8355 \tabularnewline \hline 
		$E\_Px$  & -0.1458 &$q\_D\_D\_\Pi$ & 4.7528\tabularnewline \hline 
		$E\_Py$  & -0.1458 &$q\_D\_D\_\Delta$ & 4.2950\tabularnewline \hline 
		$E\_Pz$  & -0.1458 &$I\_S\_S\_\sigma$ & 0.4\tabularnewline \hline 
		$E\_Dxy$  & -4.3034&$I\_S\_P\_\sigma$ & 0.4457  \tabularnewline \hline 
		$E\_Dyz$  & -4.3034&$I\_S\_D\_\sigma$ & -0.36819  \tabularnewline \hline 
		$E\_Dxz$  & -4.3034&$I\_P\_P\_\sigma$ & 1.5605 \tabularnewline \hline 
		$E\_Dz^{2}$  & -4.3034&$I\_P\_D\_\sigma$ & -0.2532 \tabularnewline \hline 
		$E\_Dx^{2}\_y^{2}$  & -4.3034&$I\_P\_P\_\Pi$ & -0.1348  \tabularnewline \hline 
		$V\_S\_S\_\sigma$  & -0.9588&$I\_P\_D\_\Pi$ & 0.0135 \tabularnewline \hline 
		$V\_S\_P\_\sigma$  & 1.4063&$q\_S\_S\_\sigma$ & 2.20333 \tabularnewline \hline 
		$V\_S\_D\_\sigma$  & -0.1841&$q\_S\_P\_\sigma$ & 2.6554 \tabularnewline \hline 
		$V\_P\_P\_\sigma$  & 1.4025&$q\_S\_D\_\sigma$ & 0.2495 \tabularnewline \hline 
		$V\_P\_P\_\Pi$  & -0.5730&$q\_P\_P\_\sigma$ & 1.5905 \tabularnewline \hline 
		$V\_P\_D\_\sigma$  & -0.4607&$q\_P\_P\_\Pi$ & 2.9059 \tabularnewline \hline 
		$V\_P\_D\_\Pi$  & 0.3373& $q\_P\_D\_\sigma$ & 3.8124\tabularnewline \hline 
		$V\_D\_D\_\sigma$  & -0.3709&$q\_P\_D\_\Pi$ & 3.9330 \tabularnewline \hline 
		$V\_D\_D\_\Pi$  & 0.2760& $p\_S\_S\_\sigma$ & 1.3692\tabularnewline \hline 
		$V\_D\_D\_\Delta$ & -0.0735&$p\_S\_P\_\sigma$ & 2.8794 \tabularnewline \hline 
		$I\_D\_D\_\sigma$ & -0.15&$p\_S\_D\_\sigma$ & 3.94296 \tabularnewline \hline 
		$I\_D\_D\_\Pi$ & -0.2498&$p\_P\_P\_\sigma$ & 5.5023 \tabularnewline \hline 
		$p\_P\_P\_\Pi$ & 0.536231 & $p\_P\_D\_\sigma$ & -1 \tabularnewline \hline 
		$p\_P\_D\_\Pi$ & -1 &$p\_D\_D\_\sigma$ & -0.83723 \tabularnewline \hline 
		$p\_D\_D\_\Pi$ & 0.66507 & $p\_D\_D\_\Delta$ & 4.8475 \tabularnewline \hline 
		$R_0\_inter$ & 0.25526 &$R_0\_intra$ & 0.25526 \tabularnewline \hline 
\end{tabular}\caption{TB parameters for Cu following the notation on ref~\cite{Hegde2014}}
	}
}
\label{TB_table} 
\end{table}

%\end{center}

\newpage
% Tell bibtex which bibliography style to use
\bibliographystyle{apsrev4-1}
\bibliography{GB_PR}

\end{document}